\documentclass[twocolumn,showpacs,preprintnumbers,amsmath,amssymb]{revtex4}
\usepackage{graphicx}% Include figure files
\usepackage{dcolumn}% Align table columns on decimal point
\usepackage{bm}% bold math
\begin{document}
%\draft

\title{Magnetic properties of GaMnAS from an effective Heisenberg Hamiltonian.}

\author{L.Brey$^{1}$ and G.G\'omez-Santos$^{2}$. }

\affiliation{\centerline{$^{(1)}$Instituto de Ciencia de
Materiales de Madrid (CSIC),~Cantoblanco,~28049~Madrid,~Spain.}
\centerline{$^{(2)}$Departamento de F\'{i}sica de la Materia
Condensada and Instituto Nicol\'as Cabrera,  Universidad
Aut\'onoma de Madrid~28049~Madrid,~Spain.}}

%\maketitle
\begin{abstract}

We  introduce a Heisenberg Hamiltonian for describing the magnetic
properties of GaMnAs. Electronic degrees of freedom are integrated
out leading to a pairwise interaction between Mn spins.  Monte
Carlo simulations in large systems are then possible, and reliable
values for  the Curie temperatures of diluted magnetic
semiconductors can be obtained. Comparison of   mean field and
Monte Carlo Curie temperatures shows that fluctuation effects are
important for systems with a large hole density and/or increasing
locality in the carriers-Mn coupling. We have also compared the
results obtained by using a realistic ${\bf k} \cdot {\bf p}$
model with those of a simplified parabolic two band model. In the
two band model, the existence of a spherical Fermi surface
produces the expected sign  oscillations in the coupling between
Mn spins,   magnifying  the effect of  fluctuations and  leading
to the eventual disappearance of  ferromagnetism .  In the more
realistic ${\bf k} \cdot {\bf p}$ model,  warping of the Fermi
surface diminishes the sign oscillations in the effective coupling
and, therefore, the effect of fluctuations on the critical
temperature is severely reduced. Finally, by studying the
collective magnetic excitations of the this model at zero
temperature, we analyze the stability of the fully polarized
ferromagnetic ground state.
\end{abstract}

 \maketitle
\section{Introduction}

In  recent years, one of the most studied  diluted magnetic semiconductor (DMS)
has been  Ga$_{1-x}$Mn$_x$As\cite{Matsukura,Matsukura-bis,Ohno}. Ideally, the Mn
ions substitute Ga atoms, close a $d$ shell acquiring a core spin $S$=5/2, and
give a hole to the system. Due to the experimental growth conditions, these
semiconductors have defects in the positions of  Ga and As atoms (antisite
defects), and in the location of  Mn ions (interstitials). Some of these defects
act as donors that partially compensate the holes contributed by the Mn
ions\cite{Ohno}. Therefore,   the density of holes, $p$, is  generally smaller
than the density of  Mn ions, $c$.

Experimentally, high Curie temperatures ($T_C$) are observed near an optimal
doping around  $x \sim$0.05. At this dilute concentrations,  direct interaction
between  Mn ions can be neglected. However,  Mn spins have a strong
antiferromagnetic kinetic exchange coupling, $J_{pd}$, with  hole
spins\cite{Dietlbook}.  For metallic systems, the motion of  holes mediates a
ferromagnetic interaction between the Mn ions,  leading to spontaneous
magnetization with experimental $T_C$   as high as 130K.

A first approach to the problem consists in describing the electronic system
within the virtual crystal approximation, and the thermal effects in the
mean-field approximation\cite{Dietl,Jungwirth}. In this scheme, the hole spins
feel the magnetic field created by the Mn spins which, in turn, are  equally
affected by the effective field of hole spins. Therefore,  Mn ions and  holes
become coupled by $J_{pd}$, and the low temperature phase exhibits ferromagnetism
with  Curie temperature:
\begin{equation}\label{TCMF}
k_B T _C ^{vca} = \frac{S ^2 J ^2 _{pd}}{3} \,   \chi _p ( q=0) \,
c\, \, \, ,
\end{equation}
where $\chi_p (q=0)$ is the zero wavevector paramagnetic susceptibility of the
hole gas\cite{nota1}. This expression indicates that the  Curie temperature can be
increased by raising  the density of Mn ions and/or  the number of holes. The
former possibility is limited by the tendency of the Mn ions to form clusters of
antiferromagnetically coupled Mn spins, which  do not contribute mobile holes to
the host semiconductor. The latter possibility has been explored experimentally by
different groups\cite{Potashnik1,Edmonds}. They have performed  annealing
treatments to MBE grown GaMnAs samples. In this process, the number of defects
acting as donors is reduced and, therefore, the number of holes   increases. In
this way, the Curie temperature has been raised from  75K to
110K\cite{Potashnik1}, for a Mn concentration of $x$=0.06.

The natural  question that arises concerns the possibility of further increasing
Curie temperatures.  Experimental studies of the  Mn ferromagnetic moment in
post-annealed samples show that a large fraction of  Mn spins (more than 50$\%$)
does not participate in the ferromagnetism\cite{Potashnik1}. This result opens up
the possibility of increasing  Curie temperatures by a better alignment of the Mn
spins in the zero temperature ground state of the system. Therefore, it is very
important to know whether the observed lack of magnetization saturation is due to
an extrinsic effect such as, for instance, the {\it wrong} location of some Mn
ions in the host semiconductor, or rather to intrinsic frustration  related to
the  spatial oscillatory behavior of the hole mediated interaction between  Mn
spins. The study of this last possibility provides the main motivation for this
work.

 As already mentioned, usual  mean-field calculations like that of
Eq.(\ref{TCMF})\cite{Dietl,Jungwirth} assume a virtual crystal
approximation (VCA), where  the Mn ions are replaced by an uniform
magnetic field acting on the hole spins. This approach does not
describe the individual interactions between the Mn spins and the
holes, and always predicts a fully saturated ferromagnetic ground
state (GS)  for the Mn spins. There have been several attempts to
study  the effect that thermal fluctuations and disorder have on
the value of $T_C$\cite{Schliemann,Dagotto-g,mjcdms,Berciu}, by
means of Monte Carlo (MC) simulations. By disorder we mean the
random location of the Mn ions on the FCC lattice of the host
semiconductor. These calculations have strong finite size effects
because, even though   the Mn spins are treated  classically,  the
carriers kinetic energy must be evaluated at each MC step,
requiring the diagonalization of the electron Hamiltonian.
Therefore,  MC simulations have been limited to simple electronic
models (one-orbital tight-binding Hamiltonian or two-band
parabolic model) and to small systems (less than 600 Mn ions and
60 carriers).

On the  other hand,  Schliemann  and MacDonald\cite{Schliemann1,Schliemann2}
have studied the effect that disorder and quantum fluctuations have on the zero
temperature GS, using perturbation theory and a two band model. They conclude
that long-range fluctuations make the full ferromagnetic phase unstable against a
non collinear ferromagnetic state. In the same direction and using a four band
model, Zar\'and and Jank\'o\cite{Zarand} have obtained that, due to the large
spin-orbit coupling existing in the host semiconductor,  the interaction between
Mn spins is highly anisotropic, concluding that the zero temperature GS is
intrinsically spin-disordered. Also recently, using a disordered RKKY lattice
mean field theory for a two band model, Priour {\it et al.} \cite{Priour} have
found that   Mn spins are not fully spin polarized in the zero temperature
ground state.

In this work we study the effect that disorder and thermal fluctuations have on
the value of $T_C$. We also analyze the effect that disorder and quantum
fluctuations have on the zero temperature GS of the system. To this end, we
introduce a Heisenberg-like Hamiltonian with pairwise interactions between Mn
ions for describing the magnetic properties of GaMnAs. This approach just
requires the position and spin orientation of the Mn ions, without explicit
consideration of the electronic degrees of freedom. To be precise, we propose the
following functional, \begin{equation} \label{heis} E=E_{KE}  - \sum _{I,J} J (
\mathbf{R}_{IJ}) {\bf S} _I \cdot {\bf S} _J \, \, \, \, , \end{equation} where
the coupling constant $J (\mathbf{R}_{IJ})$ is obtained using a realistic 6-band
${\bf k} \cdot {\bf p}$ Hamiltonian. The indices  $(I,J)$ run on the position of
the Mn spins, which are randomly located at sites $\mathbf{R}_I$   of the host
semiconductor FCC lattice, with $\mathbf{R}_{IJ}$=$\mathbf{R}_I$-$\mathbf{R}_J$.
We will justify the correctness of this procedure for the expected range of
parameters and, furthermore, show that $J ( \mathbf{R}_{IJ})$ can be obtained
perturbatively. As the electronic degrees of freedom are integrated inside these
coupling constants, we can perform large MC simulations for estimating the value
of $T_C$. Also, treating  Mn spins as quantum objects, we will be able  to study
collective magnetic excitations of Eq({\ref{heis}) and analyze the stability of
the zero temperature ferromagnetic GS. This will be done  by means of  an
explicit calculation of coupling constants $J (\mathbf{R}_{IJ})$ in the
spin-polarized background of carriers at zero temperature.

 The paper is organized as follows. Section II presents the microscopic model.  In
section III, we deduce  the Hamiltonian of Eq.(\ref{heis}) from
the microscopic model and justify its use. In   section IV, the
coupling constants of the pairwise Hamiltonian are obtained
perturbatively, and mean-field treatments are revised. A criterion
is introduced to quantitatively assess the effect of fluctuations
from the mere analysis of interactions. Results of MC are shown in
section V for both the two-band and six-band models. The
importance of fluctuations (thermal and disorder), their
dependence on system parameters, and their impact on the critical
temperature are  particularly considered.  In section VI, we study
our model in the zero temperature limit. We obtain the density of
states of  collective magnetic excitations and analyze the
stability of the zero temperature ferromagnetic ground state. In
section VII we finish the paper with the conclusions.

%{\it \bf{Conclusiones} We obtain in agreement with previous results that, for realistic
%values of the parameters, the two parabolic band model the combination of disorder and
%thermal fluctuations or quantum fluctuations scan suppress the ferromagnetic GS. However
%when a a more realistic six band ${\bf k} \cdot {\bf p}$  model is used a finite $T_C$
%is obtained, We discuss also the effect that a finite rarange of teh interaction betwee
%the Mn spin and teh hole spin gahve on the our results. Algo sobre Janko.}

\section{The Microscopic Model}

As explained above, we assume that  Mn ions, with a 3$d^5$ and $S$=5/2 configuration,
are randomly located in a FCC lattice. These ions donate a density $p$ of holes to
the system that, according to photoemission studies\cite{Okabayashi}, have a strong
4$p$ character that should be associated  with valence band states of the host
semiconductor. Therefore, we model  the motion of  holes in the host semiconductor,
and their interaction with the Mn spins, with the following Hamiltonian:
\begin{equation} \label{hami1}
H=H_{holes} + \sum _{I} \int d ^3 {\bf r} \, \,  {\bf S}_I \cdot
{\bf s} ({\bf r}) \, \tilde{J} ({\bf r} -{\bf R}_I) \, \, \, ,
\end{equation}
where ${\bf s} ( {\bf r} ) $ is the spin density of  carriers,
antiferromagnetically coupled to the Mn spins ${\bf
S}_I$ located at random positions ${\bf R} _I$ of the FCC lattice. The
exchange coupling $\tilde {J}({\bf r})$ has a spatial extension
that we model as a Gaussian of width $a_0$:
\begin{equation}
\tilde {J} ({\bf r})=\frac{J_{pd}}{(2 \pi a_o ^2) ^{3/2} } e ^{ -
r ^ 2 / 2 a_0 ^2} \, \, \, .  \label{jpd}
\end{equation}
$J_{pd}$ parameterizes the strength of the coupling, whereas the parameter $a_0$ is a
measure of the non-local character of this interaction. $a_0$  should be of the order
of the first neighbors distance, which corresponds to the minimum separation between
the Mn-$d$ orbitals and the GaAs-$p$ orbitals that form the hole bands of the host
semiconductor. Experimental work\cite{Okabayashi} estimates $J_{pd}
\thickapprox$60$meV$. As the Mn ions are rather diluted, we have neglected  any
direct superexchange interaction between  them.. The term
$H_{holes}$, representing the motion of holes in the semiconductor,   is described
with a   realistic six band ${\bf k } \cdot {\bf p}$ envelope function
formalism\cite{Albolfath}. In the actual magnetic semiconductors, the carrier density
is of the order of 10$^{20} cm^{-3}$ and, for these values, it seems justified to
neglect the effect of the carrier-carrier interaction. As we know\cite{Timm}  that
the  rearrangement of defects  considerably weakens  the interaction between carriers
and defects, we also neglect  any effect of disorder in the motion of carriers beyond
the mere magnetic coupling of Eq.(\ref{hami1}).

In the ${\bf k} \cdot {\bf p}$ model\cite{Cardona}, the hole
wavefunctions  are described by a
band index $n$, and a wavevector ${\bf k}$, having the form,
\begin{eqnarray}
\label{wf1} \psi _{n,{\bf k}} ({\bf r}) & = & e ^{i {\bf k} {\bf
r}} \sum _{J, m_J} \alpha _{n,{\bf k}} ^{J,m_J} u_{J,m_J} ( {\bf
r})
\\  \label {wf2}
u_{J,m_J}   ({\bf r})&  = & \frac {1} {\sqrt{N}} \sum _I
\phi_{J,m_J} ( {\bf r} - {\bf R}_I) \, \, \, \, \, ,
\end{eqnarray}
where $u_{J,m_J} ({\bf r})$ are the six $\Gamma _{4v}$  valence band wavefunctions
with $k$=0, and $\phi_{J,m_J} $ represent atomic-like orbitals with total angular
momentum $J$,  and $z$-component of the angular momentum $m_J$. The six higher energy
valence band states of GaAs correspond  to $J$=3/2 and $J$=1/2. In Eq.(\ref{wf2}),
the index $I$ runs over all the ($N$) sites of the FCC lattice, and $\sqrt{N}$ is the
normalization factor. The parameters that determine the wavefunctions, $\alpha
_{n,{\bf k}} ^{J,m_J}$, and the corresponding eigenvalues, $\epsilon _{n,{\bf k}}$,
are obtained by diagonalizing a  6$\times$6 matrix\cite{Cardona} for each wavevector.
The entries of this matrix depend on the spin-orbit coupling ($\Delta _{so}$), the
Kohn-Luttinger parameters ($\gamma _i$), and on the wavevector ${\bf k }$. We use
parameters appropriated for GaAs\cite{Cardona}, with values: $\gamma_1$=6.85,
$\gamma_2$=2.1, $\gamma_3$=2.9, and $\Delta _{so}$=0.34 eV.

In the ${\bf k} \cdot {\bf p}$ basis, the Hamiltonian of
Eq.(\ref{hami1}) has the form,
\begin{equation}
\label {hami2} H = \sum _{n{\bf k}} \epsilon _{n{\bf k}} c ^ +
_{n{\bf k}}c _{n{\bf k}} + \sum _{n{\bf k} ,n'{\bf k}'} < n{\bf k}
| V | n'{\bf k}'> c ^ + _{n{\bf k}}c _{n'{\bf k}'} \, \, \, \, ,
\end{equation}
where $ c ^ + _{n{\bf k}}$ creates a hole with quantum numbers $n$
and ${\bf {k}}$.  The second term of this equation is the
interaction between the holes and the Mn spins,
\begin{eqnarray}
\label{vnknk}& &  <  n{\bf k} |  V  |  n'{\bf k}'>  =  \frac {J
_{pd}} {\Omega}  \sum _{I}  \, e^{ -i ( {\bf k} -{\bf k}') {\bf R}
_I } \times \nonumber
\\ & &   {\bf S }_I \cdot <n',{\bf k}'|{\bf s}|n,{\bf k}> \, e ^{
-( {\bf k} -{\bf k}')^2 \frac {{a_0} ^2} {2}}
%\, (1- \delta _{{\bf k},{\bf k}'})
\end{eqnarray}
where
\begin{equation}
\label{spinhole} <n',{\bf k}'|{\bf s}|n,{\bf k}>=  \! \! \! \! \!
\sum _{Jm_J,J'm'_J} \! \! \! \! \left ( \alpha _{n,{\bf k}}
^{J,m_J}\right ) ^* \alpha _{n',{\bf k}'} ^{J',m'_J}  \, \, {\bf
s} _ {Jm_j, J'm'_J} \, \, , \nonumber
\end{equation}
the index $I$ runs over  Mn locations, ${\bf s} _ {Jm_j,
J'm'_J}$ is the matrix element of the hole spin operator in the
local angular momentum basis\cite{Albolfath}, and $\Omega$ is the
system volume. The eigen-energies of this Hamiltonian can be
written in the form
\begin{equation}
\label{def_delta} E=E_{KE}+\Delta E \, \, \, \, ,
\end{equation}
$E_{KE}$ being the energy of the carriers in absence of exchange
coupling with the Mn ions, and $\Delta E$ the variation of the
system energy due to the hole-Mn spin interaction. We take
$E_{KE}$ as our zero of energy.

\section{Pairwise interactions between  Manganese spins}

 In this section we will justify the validity of the Heisenberg-like  model of
Eq.(\ref{heis}) for the experimentally relevant range of
parameters, with coupling constants $ J (\mathbf{R}_{IJ})$
obtained perturbatively.

\subsection{ Interaction energy of a pair of Mn spins: spin and spatial
anisotropies}

Our first result is that the interaction between two Mn spins, $\mathbf{S}_1 $ and
$\mathbf{S}_2 $, is very well described by their scalar product. To show this, we
have calculated the energy of a system containing just two Mn ions: one located at
the origin with its spin pointing in the $(0,0,1)$ direction,  and the other
placed at one of the first neighbor positions $(0,a,a)/2$ of the FCC lattice
($a$=5.66$\AA$  for GaAs) with its spin pointing in the $(0,\sin\theta,\cos\theta)$
direction.   In Fig.\ref{figura_coseno} we plot this energy for different relative
orientations,  $\theta$, for a hole density $p$=0.44$nm^{-3}$, an exchange
coupling $J_{pd}$=0.06$eVnm^3$, and a spatial extension of the coupling
$a_0$=4$\AA$. We see that the interaction energy can be fitted very well by  $\cos
\theta$. Since we expect spin anisotropies to show up predominantly at short
distances, its absence for nearest neighbors makes us conclude that a spin
isotropic interaction is appropriate for arbitrarily separated Mn spins.
Therefore,
we write  the energy of a pair of Mn spins, $\mathbf{S}_1 $ and $\mathbf{S}_2$,
separated by an arbitrary vector $(i,j,k)a/2$,  in the form:
\begin{equation}\label{deltae} \Delta E = 2 \Sigma + \Delta U _{i,j,k}
\frac{\mathbf{S}_1 \cdot \mathbf{S}_2 }{S^2 }\, \, \, , \end{equation}
with
\begin{eqnarray} \label{dife2} \Delta U _{ijk} & = & \frac {1}{2} \left ( \Delta
^{(2)} _{ijk} ( \uparrow,\uparrow) - \Delta ^{(2)} _{ijk} ( \uparrow,\downarrow )
\right ) \nonumber  \\ \label{auto} \Sigma &  = & \frac {1}{4} \left ( \Delta
^{(2)} _{ijk} ( \uparrow,\uparrow) + \Delta ^{(2)} _{ijk} ( \uparrow,\downarrow)
\right ) \, \, \, , \end{eqnarray}
 where the expressions $\Delta ^{(2)} _{ijk} (
\uparrow,\uparrow)$ and $\Delta ^{(2)} _{ijk} ( \uparrow,\downarrow)$ represent
the energies of a pair of Mn spins separated by a vector $(i,j,k)a/2$, with
parallel and antiparallel spins in the $z$-direction. $\Sigma$ is a
self-interaction energy that neither depends on $(i,j,k)$ nor affects the spin
coupling.

\begin{figure}
% Requires \usepackage{graphicx}
  \includegraphics[clip,width=8cm]{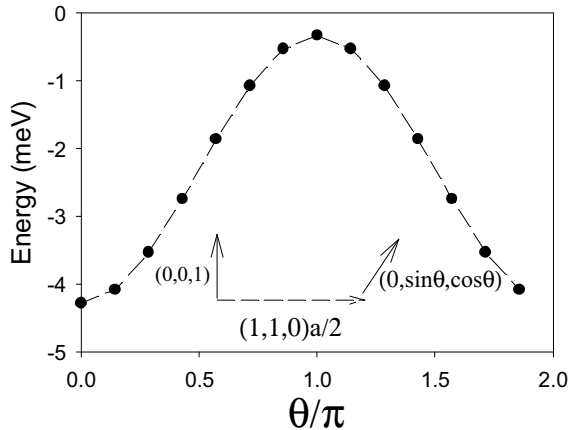} \\
  \caption{Energy of a system formed by two Mn ions. One is located at $(0,0,0)$,
  with its spin pointing in the $(0,0,1)$ direction,  the other is
  at $(0,1,1)a/2$, with its  spin oriented in the
 $(0,\sin\theta,\,cos\theta)$ direction. The calculations have been done for a hole density
 $p$=0.44$nm^{-3}$, a coupling $J_{pd}$=0.06$eV nm^{3}$, and a spatial extension of the
 interaction $a_0$=4\AA}
\label{figura_coseno}
\end{figure}

%\subsection{Spin anisotropy in the interactions.}

Recently, Zar\'and and Jank\'o\cite{Zarand} derived that the Mn-Mn
interaction in GaMnAs is highly anisotropic, using a four band
model\cite{Baldereschi}. This anisotropy is due to the
$p$-character of the top  valence bands and to the large
spin-orbit coupling existing in GaAs. In particular, they found
that the interaction between two parallel spins separated by a
vector ${\bf R}$ is different depending on whether the spins point
along  the vector ${\bf R}$ or perpendicular to it. This large
anisotropy predicted in \cite{Zarand}, seems to be in
contradiction with the results presented in Fig.
\ref{figura_coseno}, where a simple  cosine fits the interaction
very well. In order to study the spin anisotropy more carefully,
we have calculated the energy of a pair of parallel Mn spins
pointing in the $(0,\sin\theta,\cos\theta)$ direction and
separated by a vector $(0,0,a)$. In Fig. \ref{figura_dipolar} (top
panel), we plot the energy of this configuration for
$p$=0.44$nm^{-3}$, $J_{pd}$=0.06$eVnm^3$, and $a_0$=4$\AA$. We
obtain, in agreement with reference \cite{Zarand},  that the
interaction energy depends on the angle formed by the spins and
the vector joining them. Quantitatively, however,  this anisotropy
is very small: less that $10^{-4}$ times smaller than the
interaction energy and, therefore, it can be safely neglected. We
have found that this spin anisotropy becomes larger for smaller
values of the spatial extension of the interaction $a_0$, but
always remains smaller than 5$\%$ of the interaction energy, even
for the extreme case of a purely local coupling $a_0=0$. We have
checked that the discrepancy between the results of Zar\'and and
Jank\'o and ours originates in  the different models used for the
band structure. The four band chiral spherical model used in
\cite{Zarand} considers an infinite spin-orbit coupling and,
therefore,  overestimates considerably the spin anisotropy of the
interaction.

\begin{figure} % Requires \usepackage{graphicx}
\includegraphics[clip,width=8cm]{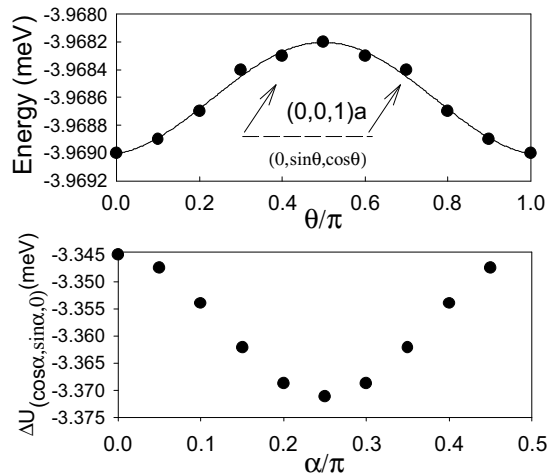} \\
\caption{Top panel: Energy
of a system formed by two Mn ions located  at $(0,0,0)$ and at $(0,0,a)$.  Mn
spins are parallel and point in the $(0,\sin\theta,\cos\theta)$ direction. The
calculations have been done for $p$=0.44$nm^{-3}$,  $J_{pd}$=0.06$eV nm^{3}$, and
$a_0$=4\AA. Bottom panel: Interaction energy of a system formed by two Mn ions
located  at $(0,0,0)$ and at $(\cos \alpha,\sin \alpha, 0)a$, for the same
parameters. Mn spins   point in the $z$-direction.}

\label{figura_dipolar}
\end{figure}

%\subsection{Spatial anisotropy in the interactions.}

Due to the $p$-character of the higher energy valence band states of GaAs,
the symmetry of the ${\bf k} \cdot {\bf p}$ Hamiltonian is cubic. Therefore,
spatial directions not related by the symmetry operations of the cubic group
are not equivalent. In order to estimate this spatial anisotropy, we have
calculated the exchange interaction energy, Eq.(\ref{dife2}), for two Mn ions
located at $(0,0,0)$ and at $(\cos \alpha,\sin \alpha, 0)a$, and with spins
oriented in the $z$-direction. Using this geometry, we avoid the effects
associated with the spin anisotropy discussed before, leaving alone the
spatial anisotropy.  In Fig.\ref{figura_dipolar} (bottom panel) we plot the interaction
energy as a function of the angle $\alpha$, for $p$=0.44$nm^{-3}$,
$J_{pd}$=0.06$eV nm^{3}$, and  $a_0$=4\AA. We observe that the interaction
energy  depends very weekly on the the orientation of the vector that joins the
Mn spins. As in the case of spin anisotropy, the effect of the spatial
anisotropy is also larger for smaller values of $a_0$, but always remains
smaller than $5\%$. In conclusion, an  expression like Eq.(\ref{heis}), with a
coupling constant dependent on distance but not on orientation, provides an
adequate description for the interaction between two Mn spins.

%\begin{figure}
% Requires \usepackage{graphicx}
%  \includegraphics[clip,width=8cm]{figura_isotropia.eps} \\
%  \caption{Interaction energy of a system formed by two Mn ions located  at $(0,0,0)$ and
%  at $(\cos \alpha,\sin \alpha, 0)a$. The spins of the Mn ions  point in the
% $z$-direction. The calculations have been done for
% $p$=0.44$nm^{-3}$,  $J_{pd}$=0.06$eV nm^{3}$ and  $a_0$=4\AA}
%\label{figura_isotropia}
%\end{figure}

%\subsection{Energy as a sum of pair interactions.}

\subsection{Virial expansion and perturbation theory}

 Previous MC simulations\cite{mjcdms} have shown that the Curie temperature of
DMS increases linearly with  $x$ for low Mn concentration. This linearity
strongly suggests that  a virial-like approach, where the Mn interaction is
expressed as a sum of pairwise terms, should provide a good approximation in
this low concentration regime.

In order to test the validity of this approach, we have computed the energy of
a three Mn ions system, comparing the result with the energy of three systems
containing only two Mn ions. We place three Mn ions at close positions
$(0,0,0)$, $(1,0,0)a$, and $(0,1,0)a$, and call $\Delta ^{(3)}$   the exact
energy of this three-body system.   Assuming only pairwise interactions, the
energy to be associated to this system can be written as (see
Eq.(\ref{dife2}):
\begin{equation}
\label{3apares} \Delta ^{(3)} _{pairs} =  3 \Sigma + \Delta
U_{020} +\Delta U _{200}+\Delta U _{220} \, \, \, \,  \end{equation}
 The
comparison between the exact and pairwise approximated energies for the
parameters  $p$=0.44$nm^{-3}$, $J_{pd}$=0.06$eVnm^{-3}$, and $a_0$=4\AA,
 provides the following results:
\begin{equation}
 \Delta ^{(3)}=-5.043 meV, \;\;\; \Delta ^{(3)} _{pair}=-5.031meV
 \end{equation}
  From these numbers, we conclude that writing the energy of the system as a sum
of pair interactions is indeed a very good approximation for DMS.

The interaction energies  presented in the previous subsections
were obtained by solving the Hamiltonian (\ref{hami1}). From these
calculations we have justified the use of a Heisenberg-like
Hamiltonian for describing the magnetic properties of GaMnAs. In
this approach, the coupling constants $J(\mathbf{R}_{IJ})$ have
been obtained by solving Eq.(\ref{hami1}) for different distances
between  Mn ions, and for parallel and antiparallel orientation of
their spins. The exact solution of the Hamiltonian (\ref{hami1}),
even for only two Mn ions as  in the previous subsections,
requires the diagonalization of very large complex matrices
(larger than 3000$\times$3000),  posing a severe computational
problem.

 To bypass this difficulty, we have resorted to perturbation theory  for the
calculation of  coupling constants. We know that, for small values of the
exchange coupling, a second order perturbation theory treatment of $V$
(Eq.(\ref{vnknk})) should be  valid, leading to   interaction energies
proportional to $J_{pd}^2$. In order to check the validity of a perturbative
treatment, we have calculated  the interaction energy of a system formed by two
Mn spins located at positions $(0,0,0)$ and $(1,1,0)a/2$, for the parameters
$p$=0.44$nm^{-3}$, and $a_0$=4\AA.  The results, plotted in
Fig.(\ref{figura_cuadratico}), confirm that the interaction energy remains
quadratic until values of the exchange coupling of the  order $J_{pd} \sim
$100$meVnm^{-3}$. Therefore, we conclude that the use of perturbation theory
for computing the interaction energies is appropriated, enormously simplifying
the computational effort. It is worth mentioning here  that, although  previous
MC simulations \cite{Schliemann,mjcdms} have found deviations form the
quadratic dependence of the Curie temperature on $J_{pd}$, however, these
deviations appear at values of $J_{pd}$ larger than 0.1eV$nm^{-3}$.

It is interesting to note that the perturbative regime, with its associated small
values of  coupling constants,  fits nicely within the virial expansion approach.
Three-body interactions will only appear  to order $J_{pd}^3$, making us expect our
pair interaction approximation to be valid even for not so low impurity
concentrations, provided the coupling constant remains small.

\begin{figure}
% Requires \usepackage{graphicx}
  \includegraphics[clip,width=8cm]{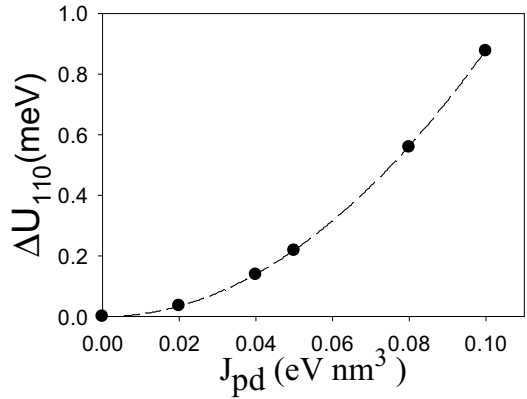} \\
  \caption{Interaction energy as function of $J_{pd}$ of a system formed by two Mn ions located  at $(0,0,0)$ and
  at $(1,1,0)a/2$.  The calculations have been done for
 $p$=0.44$nm^{-3}$ and  $a_0$=4\AA}
\label{figura_cuadratico}
\end{figure}

\section{Heisenberg Hamiltonian. Mean-Field Treatments and Fluctuations}
The calculations presented in the previous section justify
the description of magnetic properties of GaMnAs in terms of
a two body, spin isotropic, Heisenberg-like Hamiltonian:
\begin{equation}
\label{heis2}
  H= -\sum _{I,J} J(\mathbf{R}_{IJ}) {\bf S}  _I \cdot {\bf S} _J
\, \, \, \, ,
\end{equation}
with exchange coupling  given by
\begin{equation}
\label{jij} J({\bf R}_{IJ}) =  \frac{1}{N} \sum _{ {\bf q}}
j({\bf q}) e^{i{\bf q}{\bf R}_{IJ} } \, \, \, .
\end{equation}
%where $\Omega$ represents the system's volume, and ${\bf R}_{IJ}$ is the vector joining the locations of the spins $ {\bf S} _I$ and
%$ {\bf S} _J$.
The Fourier transformed coupling, $j({\bf q})$, is given by
\begin{equation}
\label{jchiq}
j ({\bf q})= J_{pd} ^2 \, \frac{N}{\Omega } \, \chi
_p ( {\bf q}) \, e^{-q^2 a_0 ^2} \, \,
\end{equation}
with the paramagnetic susceptibility obtained in the standard perturbative manner
 form the original Hamiltonian of Eq. \ref{hami1}:
\begin{eqnarray}
\label{jq}
\chi ({\bf q}) & = &    \frac{1}{\Omega} \sum
_{n',n,{\bf k}} \frac { n_F (\epsilon _{n,{\bf k}})-n_F(\epsilon
_{n',{\bf k}+ {\bf q}})} {\epsilon _{n',{\bf k}+ {\bf q}}-\epsilon
_{n,{\bf k}}} \nonumber \\ & & |<n',{\bf k}+ {\bf q}|s_z|n,{\bf
k}> |^2 \, \, \, .
\end{eqnarray}

 As explained in the previous section, we will consider a space-isotropic
susceptibility $ \chi (q) $, obtained by angular averaging the true susceptibility.
This leads  to a function $j ( q)$ depending only on the modulus of $\mathbf{q}$, from
which real space couplings are easily extracted.

Our main concern will be the obtention of critical temperatures
from MC simulations. Nevertheless, we will also compare MC
calculations with the predictions of mean-field approaches. We
believe this is interesting for, given the usual computer
limitations of MC calculations, this comparison allows us to gauge
the merits and shortcoming of standard mean-field techniques. In
addition, we will show that simple considerations concerning the
effect of fluctuations, allow us to estimate how far the
mean-field temperature is from the real one. Therefore, we devote
the rest of this section of this analysis.

 In the standard mean-field approach, the critical temperature can be read-off from the
 effective field experienced by a magnetic impurity. In our site disordered system, this can
 be written as follows:
\begin{equation}
k T _C ^{mf} = \frac{S ^2}{3}  < \sum _{I\neq 0} x_i J({\bf R}_{I}) > =
 \frac{S ^2}{3} x  \sum _{I\neq 0} J({\bf R}_{I})\, \,
\, \, \, ,
\end{equation}
 Where the sum runs over all sites of the FCC lattice and the average is taken over
 disorder configuration. The latter is  characterized by the random variable $x_i  $, which marks the
 presence ($x_i = 1$) or absence ($x_i = 0 $) of impurity at site $i$, with average
 $<x_i>=x$.

  The sum can be evaluated in Fourier space as follows
\begin{eqnarray}
k_B T _C ^{mf} & = & \frac{S^2}{3}\, x \, \left (\sum _{i} J({\bf
R}_{I}) -J ({\bf
R}=0)   \right ) \nonumber \\
& = &\frac{S^2}{3}\, x \, \left (J_{pd} ^2 \frac{N}{\Omega } \chi _p ( {\bf
q}\rightarrow 0) - \frac{1}{N } \sum  _{\bf q} j ({\bf q}) \right )
\nonumber \\
&=&k_B  T _C ^{vca} \left ( 1- \frac{1}{N }\sum  _{\bf
q}\frac{\chi _p (  q)e^{-q^2 a_0 ^2}}{\chi _p ( q \rightarrow 0)}
\right ) \, \, \, .
\end{eqnarray}

%with $T _C ^{vca}$ given by
%\begin{equation}
%T _C ^{vca} =\frac{S^2}{3} \, x \, J_{pd} ^2 \frac{N}{\Omega} \chi _p (q \rightarrow
%0)
%\end{equation}

 Notice that the self-interaction must be explicitly removed. If it is not, one
ends up with a different mean-field approximation which we have previously termed
the virtual crystal approximation $T _C ^{vca}$. Looking at  the Heisenberg
Hamiltonian of Eq. \ref{heis2}, it is clear from a conceptual point of view that
the {\em genuine} mean-field approximation requires this removal of the
self-interaction term, even if both temperatures turn out to be  similar in the
physical region of parameters . It is this suppression of self-interaction what
accounts for the reduction, and even the disappearance, of the critical temperature
with increasing carrier concentration, owing to the oscillatory nature of the
exchange coupling. While  $T _C ^{mf}$ accounts for this effect, the VCA approach
is blind to this oscillations, always predicting a finite transition temperature.

It is well known that mean-field approximations tend to overestimate the
transition      temperature due to the neglect of fluctuations. Nevertheless,
the effect of the neglected fluctuations, and their influence in the transition
temperature, can be estimated by means of a consistency criterion, similar in
spirit to the well-known Ginsburg criterion
(see any text on critical phenomena, for instance, Ref.\cite{cardy96}),
as we now explain. The mean-field approach assumes independent spins under the
influence of a molecular field determined self-consistently. At $T=T_C ^{mf} $,
the average molecular field is zero, and so is the magnetization. Yet, even within
this mean-field scenario, there are fluctuations around the zero average field. In
the present case where the interaction can extend over large distances for small
carrier concentration, this molecular field will be the sum of many random
variables, therefore a Gaussian distribution is expected for it. At the critical
temperature, we can write the following expression for the distribution of
molecular fields along the z axis
\begin{equation}
{\cal P}(h_{mol}) =\frac{1}{(2 \pi \Delta h_{mol})^{1/2}} \exp
\left(- \frac{h_{mol}^2}{2 (\Delta h_{mol})^2 }\right),
\end{equation}
with a dispersion in local fields given by
\begin{equation}
(\Delta h_{mol})^2 = \frac{x S^2}{3} \sum_{{I\neq 0}} ( J({\bf R}_I) )^2,
\label{deltahmol}
\end{equation}
where the lattice sum can be evaluated in Fourier space as
\begin{equation}
\sum_{{I\neq 0}} ( J({\bf R}_I) )^2 = \frac{1}{N} \sum_{{\bf q}} |J({\bf q})|^2
- \left( \frac{1}{N} \sum_{{\bf q}} J({\bf q})\right) ^2.
\end{equation}
Notice that both thermal and positional disorder contribute to the expression
of Eq. \ref{deltahmol}.

This means that, even though the average (spontaneous) magnetization is zero at
the nominal critical temperature
$T_C ^{mf} $, there will be a distribution of local magnetizations, with a
dispersion given by
\begin{equation} <m^2>_{T_C ^{mf}} = \int d h_{mol} \; {\cal P}(h_{mol}) \; (S \tilde{{\cal
M}}(\tilde{h}_{mol}))^2
\end{equation}
where
\begin{equation}
{\tilde{\cal M}}(\tilde{h}_{mol}) = \left(\frac{1}{\tanh(\tilde{h}_{mol})} -
\frac{1}{\tilde{h}_{mol}}\right)
\end{equation}
is the normalized magnetization of a isolated impurity in
the presence of the (dimensionless) molecular field $\tilde{h}_{mol} =
S h_{mol}/(k T_C ^{mf})$.

This provides us with a natural consistency criterion for the validity of mean-field
results. We can say, for instance, that we will trust the mean-field results for
temperatures  $ T \leq T^{*} < T _C ^{mf}$ such that  the average, spontaneous, mean
field magnetization $ m(T^{*})$ is much larger than the "incertitude"  of the
magnetization at $ T _C ^{mf}$ due to fluctuations (several variants of this criterion
can be envisaged). To be specific, we will define a temperature $ T^{*}$   such that
\begin{equation}
 \left( m(T^{*}) \right) ^2 = {\cal G} <m^2>_{T_C ^{mf}}
 \label{ginsburg}
\end{equation}
where  $ {\cal G} $ is a "large", dimensionless parameter, and the  mean-field
magnetization is the well-known implicit solution of  $ m= S \tilde{{\cal
M}}(3 \frac{T}{T_C ^{mf}} \frac{m}{S}) $. We have chosen the value of ${\cal G}$ by
looking at the simplest Heisenberg model:  classical spins in a cubic lattice with nearest neighbor
couplings. In this simpler {\em reference} model, the value of ${\cal G} $ is obtained
demanding that,
if the previous scheme is
applied, the $T^{*}$ so obtained coincides with the MC determined temperature for
that model\cite{Lau}. The actual value is around ${\cal G}=8.3$. Irrespective of the precise numerical value of the coefficient
${\cal G}$, it is clear that the $T^{*}$ so obtained is a direct measure of the
effect of fluctuations and, therefore, of the corrections to the mean-field
temperatures. We will term this  $T^*$ the fluctuations-corrected critical
temperature.

\section{Results: Exchange Interactions, Curie Temperatures, and Magnetization}
\subsection{Two-band model}
\begin{figure}
% Requires \usepackage{graphicx}
 % \includegraphics[clip,width=8cm]{figura_jq2b.eps} \\
  \includegraphics[clip,width=8cm]{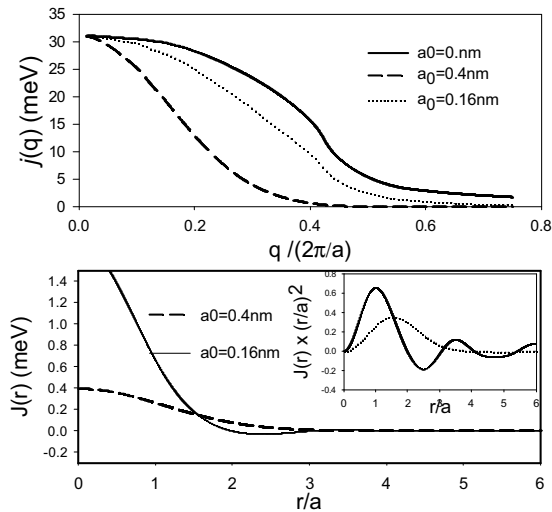} \\
  \caption{ Top panel: Fourier transform of the interaction energy as function of $q$ for a two-band model with
  $m^*$=0.5,
 $p$=0.44$nm^{-3}$, $J_{pd}$=0.06$meVnm^{3}$, and two different values of the exchange coupling spatial extension
  $a_0$=0 and
 $a_0$=4\AA.
Bottom panel: Real space exchange coupling as function of $r$ for a two-band model with
  the same parameters.}
\label{figura_jq2b}
\end{figure}

As a first example, we apply or method to the simplest case of a
two-band model. In this model, the wavefunctions are plane waves
of momentum ${\bf k}$ with a dispersion relation $\hbar ^2 k ^2
/(2m^*)$,  $m^*$ being the average effective mass of the holes. In
Fig.({\ref{figura_jq2b}) (top panel)  we show the Fourier
transform  of the exchange interaction for $p$=0.44$ nm^{-3}$,
$J_{pd}$=0.06$ meV nm^{3}$, $m^*$=0.5, and different values  of
$a_0$: 0, 1.6\AA,  and 4\AA. For $a_0$=0, $j(q)$ is proportional
to the paramagnetic susceptibility, which coincides with the
Lindhard function\cite{Pines_Nozieres} and exhibits the well known
anomaly in the second derivative at $2\,k_F$.  For $a_0$=0, the
paramagnetic susceptibility of the carriers decays very slowly as
a function of $q$, and the value of the coupling $j(q)$ is
significant even for values of $q$ near the zone  boundary
$2\pi/a$. This result is unlikely as the effective mass
approximation and the ${\bf k} \cdot {\bf p}$ method are just
applicable for describing the semiconductor band structure
near the center of the Brillouin zone. Fortunately, this problem
is solved when we introduce a finite spatial extent ($a_0$) for
the coupling between Mn and carriers. In the paper we perform
calculations for $a_0$= 1.6\AA \ and 4\AA. For $a_0$=1.6\AA,   the
interaction decays rather slowly in reciprocal space, although its
value at the zone boundary is practically zero,
Fig.(\ref{figura_jq2b}). For $a_0$=4\AA, the interaction decays
rather fast in  reciprocal space, being nearly zero at
half the Brillouin zone. The value $a_0$=4\AA \ corresponds to the
first neighbors distance in the FCC lattice with $a$=5.66\AA.

In Fig.(\ref{figura_jq2b}) (bottom panel) we show
the real space coupling constant as a
function of $r/a$ for the same parameters. For $a_0$=0, this is nothing but the
Ruderman-Kittel-Kasuya-Yosida (RKKY) interaction:
\begin{equation}\label{rkky_int}
    J_{two \, \, band} (r)   \sim \frac{2k_F r\cos (2k_F r)-\sin (2k_F r  ) }{(2k_F r )^4
    }\, \, \, .
\end{equation}
This function decays as $r^3$ and  oscillates in sign with the
distance. This oscillatory behavior is a signature of  the $2k_F$ anomaly
that occurs in the paramagnetic susceptibility.  For finite values
of $a_0$,  the $2k_F$ anomaly in $j(q)$ is softened, and the
oscillatory behavior of the real space coupling, $J(r)$, is
damped, although for $a_0$=1.6\AA \ it remains  notable. It is
important to remark the rather long-range character of the interaction
between Mn spins. Even for the case of $a_0$=4\AA, the
coupling is significant up to distances three times larger  than
the FCC lattice parameter. Note that, in the FCC lattice, the
number of neighbors for a distance cutoff of 3$a$ is  around
500. Therefore,  in the Heisenberg Hamiltonian (Eq.\ref{heis2}),
the number of neighbors to be taken into account is of the
order of $x \times$500, at least.

%\begin{figure}
% Requires \usepackage{graphicx}
%  \includegraphics[clip,width=8cm]{figura_jr2b.eps} \\
%  \caption{Real space exchange coupling as function of $r$ for a two-band model with
%  the same parameters than in Fig.(\ref{figura_jq2b}).}
%\label{figura_jr2b}
%\end{figure}

Once the real space exchange coupling constants are known, we perform classical MC
simulations on the orientation of the Mn spins. We use a FCC super cell of volume
$N^3 \frac{a^3}{4}$ with $N=30$, including more than 1200 Mn ions. For this system
size, we have checked that the results are free of size effects, and that the
disorder in the Mn ion positions is self averaged. Due to the long-range character
of the interaction, we have to include the interaction of each Mn spin with its
first 150 neighbors in the MC simulation. This corresponds to a real space cutoff
bigger than 5$a$. For distances larger than 5$a$, the interaction  can be neglected,
as shown in  Fig.  \ref{figura_jq2b}.

The most notable result is the absence of a finite Curie
temperature in the MC simulations for small values of $a_0$ and
large values of the hole density. We have checked the absence of
spontaneous magnetization for $a_0$=1.6, and carrier densities
above $p=0.44 nm^{-3} $.
%(for these small values of $a_o$, the
%range of the interaction increases beyond our MC capabilities for
%densities below $p=0.44 nm^{-3}$).
For a density  $p$=0.22$nm^{-3}$, we obtain a  Curie temperature
of 14K, considerably smaller than the mean field value. It is
interesting to realize that the fluctuation analysis described
previously fits nicely with these MC results. In Fig. \ref{2b-1.6},
we plot the predictions of the mean-field treatments, along with
the expected corrections from the neglect of fluctuations. We see
that, for $p\sim 0.44 nm^{-3}$, the fluctuations have grown to the
point that no finite temperature exists meeting the consistency
criterion described in Eq. \ref{ginsburg}, in agreement with the
absence of magnetization observed in MC simulations. Notice that,
for lower carrier concentration, a finite magnetization
re-emerges. In fact, the mean-field temperatures $ T_C ^{mf}$, $
T_C ^{VCA}$, and the fluctuations-corrected temperature $ T^*$,
merge in the very low density  limit, as shown in Fig.
\ref{2b-1.6}. This is not surprising, for a vanishing $k_F$
implies long-ranged interactions where the neglect of fluctuations
becomes increasingly irrelevant, leading to  the asymptotic
correctness of mean-field results.

\begin{figure}
% Requires \usepackage{graphicx}
  \includegraphics[clip,width=8cm]{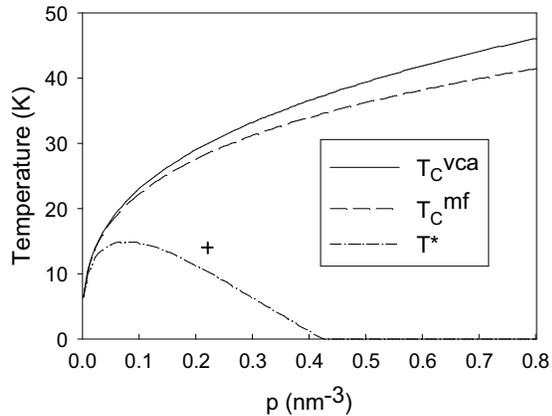} \\
  \caption{VCA mean-field Curie temperature ($T _C ^{vca}$),
  self-interaction corrected Curie temperature ($T _C ^{mf}$), and fluctuations-corrected
  Curie temperature ($T^*$) for $a_0$=1.6\AA $\;$ and a Mn concentration
$x$=0.05. The cross represents the MC temperature for $p$=0.22$nm^{-3}$. No
magnetization is observed above $p$=0.44$nm^{-3}$ in the MC simulations.}
\label{2b-1.6}
\end{figure}

\begin{figure}
% Requires \usepackage{graphicx}
%  \includegraphics[clip,width=8cm]{figura_mag2b.eps} \\
%  \vspace{1cm}
%  \caption{caca}
%  \includegraphics[width=8cm]{2b-4.eps}
   \includegraphics[width=8cm]{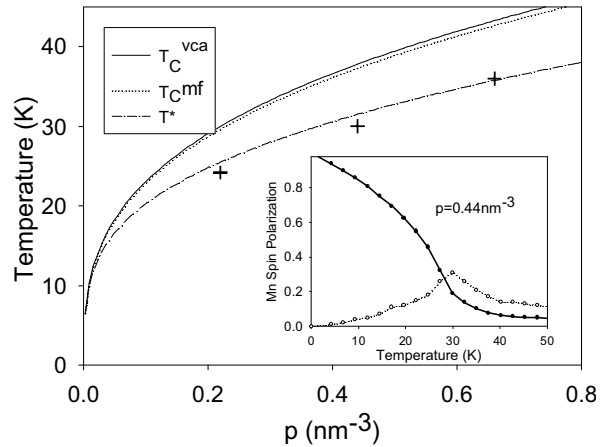}
  \caption{
VCA mean-field Curie temperature ($T _C^{vca}$), self-interaction
corrected Curie temperature ($T _C ^{mf}$), and
fluctuations-corrected Curie temperature ($T^*$) as a function of
carrier density, for a two-band model with $m^*$=0.5,
$J_{pd}$=0.06$meVnm^{3}$, $a_0$=4\AA,  and a Mn concentration
$x$=0.05. MC Curie temperatures corresponding to different hole
densities are  represented by crosses. Inset: Close dots represent
the temperature dependence of the Mn spin polarization as obtained
from MC simulations for $p$=0.44$nm^{-3}$. Open dots correspond to
Mn spin polarization fluctuations, that help to estimate the value
of the Curie temperature. } \label{figura_mag2b}
\end{figure}

Increasing the spatial extension of the coupling between
impurities and carriers to $a_0$=4\AA, the RKKY oscillations are
damped and  a low temperature ferromagnetic GS appears in MC
calculations.  In the inset of Fig. (\ref{figura_mag2b}),  we plot
the temperature dependence of the Mn spin polarization as obtained
from MC simulations  for a two-band model with $m^*$=0.5,
$p$=0.44$nm^{-3}$, $J_{pd}$=0.06$meVnm^{3}$, $a_0$=4\AA, and a Mn
concentration $x$=0.05. In   Fig. (\ref{figura_mag2b}), we plot the
MC Curie temperatures for different hole densities, along with the
fluctuations-corrected $T^*$ and  the mean-field results, both
with and without self-interaction corrections. The observed MC
temperatures are consistent with the fluctuations-corrected $T^*$.
From this analysis, it is clear that Curie temperatures approach
the mean-field results with increasing non-locality (larger values
of $a_0$) in the coupling between impurities and carriers.

\subsection{ $ {\bf k} \cdot {\bf p} $ model}

\begin{figure}
% Requires \usepackage{graphicx}
  \includegraphics[clip,width=8cm]{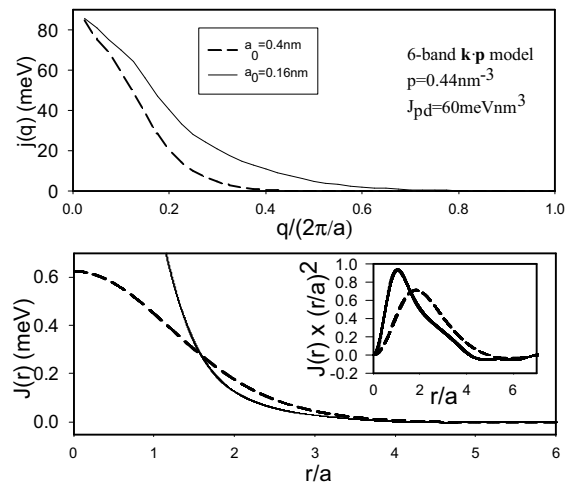} \\
  \caption{Top panel: Fourier transform of the interaction energy as function of $q$ for the $\mathbf{k} \cdot \mathbf{p}$
  model with
 $p$=0.44$nm^{-3}$, $J_{pd}$=0.06$meVnm^{3}$, and two different values of the exchange coupling spatial extension
  $a_0$=3\AA and
 $a_0$=4\AA.
 Bottom panel: Real space  coupling as function of $r$ for the same parameters.}
\label{figura_jqkp}
\end{figure}

 The $ {\bf k} \cdot {\bf
p} $ model describes  the valence band of the semiconductors in a
realistic way.  The most characteristic features of the GaAs
valence bands are the anisotropy in reciprocal space and the
strong coupling between the heavy and light hole bands. These
effects substantially alter  the constant energy surfaces of the
holes states, which become non spherical, with a warped shape. The
lack of a well-defined modulus of the Fermi wavevector softens the
$2k_F$ anomaly in the average paramagnetic susceptibility as a
function of $q$. This is clear in Fig(\ref{figura_jqkp}) (top
panel), where we plot the Fourier transform of the interaction
energy for the same parameters as in Fig.(\ref{figura_jq2b}), for
$a_0$=1.6\AA$\,$ and $a_0$=4\AA.   In the bottom panel of
Fig.(\ref{figura_jqkp}), we plot the real space energy coupling
$J(r)$. It decays   almost continuously to zero with a very
weak oscillation, as expected from the absence of strong anomalies
in $j(q)$.  As in the two band model, we obtain that the
interaction between Mn ions extends several lattice sites, and the
quantity of merit, $J(r) r^2$ (see inset of
Fig.(\ref{figura_jqkp})), has a maximum near $r=2a$, for both
values of $a_0$=1.6\AA \ and $a_0$=4\AA.

\begin{figure}
% Requires \usepackage{graphicx}
  \includegraphics[clip,width=8cm]{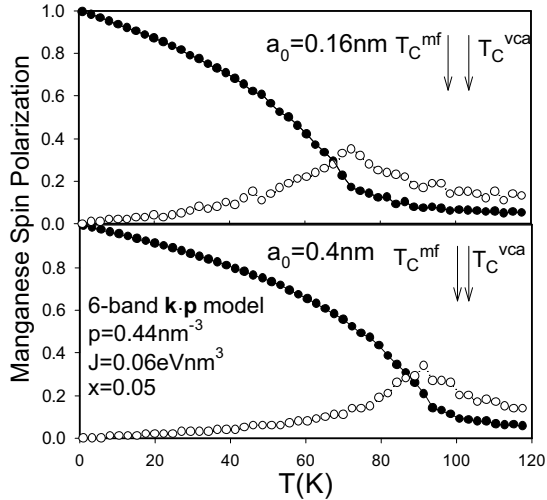} \\
 \caption{Close dots represent  the temperature dependence of the
  Mn spin polarization as obtained from MC simulations
 for the $\mathbf{k} \cdot \mathbf{p}$
with  $p$=0.44$nm^{-3}$, $J_{pd}$=0.06$meVnm^{3}$, a Mn
concentration $x$=0.05, and two different values of $a_0$, 1.6\AA \
and 4\AA. Open dots correspond to Mn spin polarization
fluctuations, that help to estimate the value of the Curie
temperature.}

 \label{figura_magkp}
\end{figure}

\begin{figure}
% Requires \usepackage{graphicx}
  \includegraphics[clip,width=8cm]{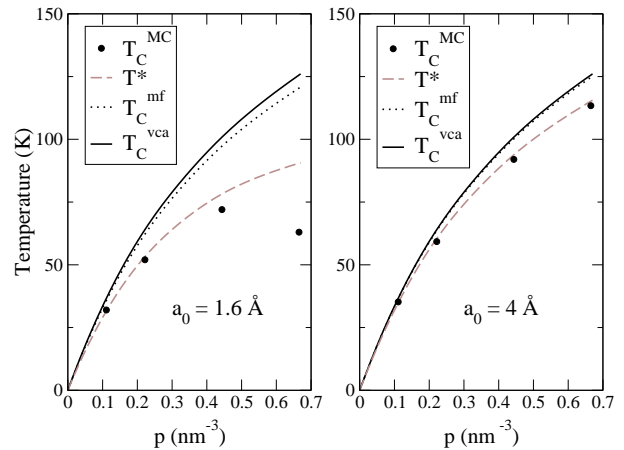} \\
 \caption{$\mathbf{k} \cdot \mathbf{p}$ model Curie  temperatures from MC calculations
 as a function of hole density, compared with results of mean-field
 temperatures: $T _C ^{vca}$,$T _C ^{mf}$,
 and the fluctuations-corrected critical temperature $T^*$, for Mn concentration
 $x$=0.05}
\label{figura_dfkp}
\end{figure}

We have performed MC simulation with the coupling constants showed in
Fig(\ref{figura_jqkp}) and a concentration of Mn ions of 5$\%$. The details of the
simulation are the same as in the two band case. Representative results for the Mn
spin polarization  as a function of temperature are shown in
Fig.(\ref{figura_magkp}). In Fig. (\ref{figura_dfkp}), we exhibit the MC Curie
temperatures along with the mean-field  and fluctuation-corrected results. Comparing
with the corresponding results for the two-band case, the six-band model shows an
overall increase of transition temperatures. In addition, the MC temperatures are
closer to the mean-field results than in the two-band case, for the same parameters.
It is interesting to note that, again, the fluctuation-corrected temperature ($T^*$)
offers a fair estimation of the MC temperature. In any case, the combined effect of
thermal and disorder fluctuation are much less severe than in the two-band case. This
is evident, for instance, in the curves for $a_0$=4 \AA, where  the VCA transition
temperature  $T_C ^{vca}$ already provides a good estimation of the MC $T_C$. This is
consistent with the fact that interactions in this six-band case, while extending
several lattice sites, do not manifest the violent sign oscillations present in the
two-band calculation. As in the two-band calculation, mean-field and exact (MC)
temperatures tend to merge at low carrier density, as expected from the increasing
range of the exchange coupling between Mn ions. Similarly, increasing the spatial
extent of the interaction between Mn ions and carriers ($a_0$) softens the effect of
fluctuations, leading to a better agreement between mean-field and MC
temperatures.

\section{ Collective excitations of the $T$=0 Ground State}

In this section, we study the low energy collective magnetic
excitations of DMS at zero temperature. We assume that, at  $T$=0,
 Mn spins are fully polarized, and the spin polarization of the
carriers, $\xi$, is  obtained by solving the carriers Schrodinger
equation in the presence of the uniform magnetic field created by
the Mn ions, $S J_{pd}c$. The low energy
collective magnetic excitations of the system are obtained from
the Heisenberg Hamiltonian,
 \begin{equation}
  H_{\xi} = E _{KE} (\xi)- \sum _{I,J} J _{\xi} ({\bf R} _{IJ} )
  \,
  {\bf S} _I \cdot {\bf S} _J \, \, \, ,
  \label{heis_pol}
  \end{equation}
where $E_{KE}(\xi) $ is the kinetic energy of the carriers
evaluated at the hole spin polarization $\xi$. This Hamiltonian
describes small oscillations of the Mn spins from the fully
polarized state, that we choose to point in the $z$-direction. The
Fourier transform of the coupling constants $ J _{\xi} ({\bf R}
_{IJ} )$ is proportional to the  transverse response function of
the polarized hole gas $ \chi ^{\perp} ( q, \xi)$,
\begin{equation}
j_{\xi} (q) = J ^2 _{pd} \frac{N}{\Omega} \chi ^{\perp} ( q, \xi)
e ^{-q ^2 a_0 ^2 } \, \label{jqxi}
\end{equation}
In writing Eq.(\ref{heis_pol}) and Eq.({\ref{jqxi}) we  assume, as
we have justified in previous sections,  that the Mn interaction
energy can be expressed as a sum of Mn spin pair interactions,
which only depends on the relative angle formed by the Mn spins and
on the distance separating  them. We have also neglected the
magnetic anisotropy energy that we know is very
small{\cite{Dietl,Jungwirth}.

The low energy collective excitation are obtained by solving the
equation of motion of the Mn spins,
\begin{equation}
-i \hbar \frac{\partial S _I ^- }{\partial t } = [ H_{\xi}, S _ I
^ - ] = - \sum _J J _{\xi} ({\bf R} _{I,J} ) \left ( S _i ^z S _J
^- - S _ J ^z S _ I ^-) \right ) \,  \, . \label{eq_mot}
\end{equation}
Near $T$=0, the Mn spins are fully polarized, we replace $S^z$ by
$S$ and the equations may be linearized.

\subsection{Spin waves in the VCA}
\begin{figure}
% Requires \usepackage{graphicx}
  \includegraphics[clip,width=8cm]{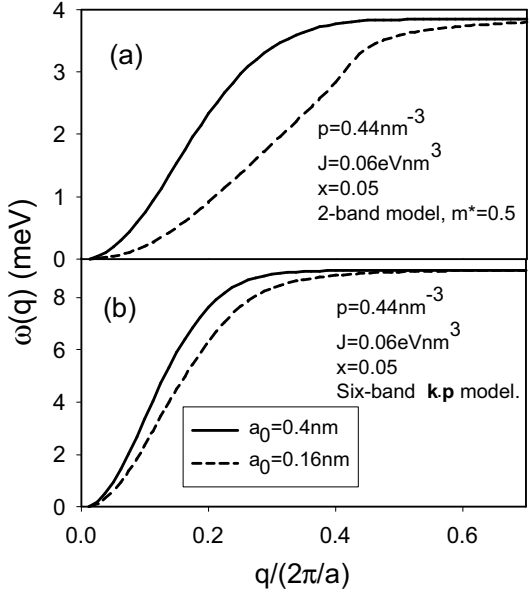} \\
 \caption{Spin wave dispersion, obtained in the virtual crystal approximation, for the two-band model and for the
  $\mathbf{k} \cdot \mathbf{p}$ model.
   The results correspond to
  $p$=0.44$nm^{-3}$, $J_{pd}$=0.06$meVnm^{3}$, a Mn concentration
$x$=0.05 and two different values of $a_0$, 1.6\AA and 4\AA. The
carriers spin polarization is $\xi$=0.69 for the two band model
and $\xi$=0.69 for the $\mathbf{k} \cdot \mathbf{p}$ model. }
 \label{figura_sw_mf}
\end{figure}
In the VCA,  all sites are equivalent and, in Eq.({\ref{eq_mot}),
the sum over the Mn positions is replaced by a sum over all
lattice position times the Mn concentration $x$. In this
approximation the collective excitations are spin waves with
momentum ${\bf q}$ and dispersion relation
\begin{equation}
\omega ({\bf q}) = x S \left [ J _{\xi} ( q=0) - J _ {\xi} (q)
\right ] \label{omegaq} \, \, \, .
\end{equation}
In Fig.(\ref{figura_sw_mf}) we plot the spin wave dispersion for
the two-band model and for the six-band ${\bf k } \cdot {\bf p}$
model with typical values of the parameters and for both $a_0$=4\AA \ and
$a_0$=1.6\AA. The spin waves are gapless as the model Hamiltonian
(\ref{heis_pol}) is invariant under rotations. At small values of
the wavevector, the spin waves disperse quadratically and the
expression  $\omega (q)= \rho_s q ^2 /(2 \pi / a)$ defines the
spin stiffness $\rho _s$. The spin waves are harder in the
six-band ${\bf k} \cdot {\bf p}$ model ($\rho_s$=242meV and 370meV,
for $a_0$= 1.6\AA \ and $a_0$=4\AA,  respectively ) than in the two
band model ($\rho _s$=18meV and 76meV, for $a_0$= 1.6\AA \ and
$a_0$=4\AA,  respectively). This is consistent  with the fact that
the Curie temperature is larger in the ${\bf k} \cdot {\bf p}$
model than in the two band model. Also we find that the stiffness
increases as $a_0$ increases. These numerical results for the
stiffness  agree with those obtained by Konig {\it et al.} using a
formalism that treat the Mn spins as non-interacting
bosons\cite{Konig}.
%\begin{figure}
%% Requires \usepackage{graphicx}
%  \includegraphics[clip,width=8cm]{figura_dos_mf.eps}
% \caption{ Low energy excitations density of states, as obtained in the virtual crystal approximation,
% for the two-band model and for the
%  $\mathbf{k} \cdot \mathbf{p}$ model.
%  The parameters used are the same than in
%  Fig.(\ref{figura_sw_mf}).}
% \label{figura_dos_mf}
%\end{figure}
%
%From the dispersion of the spin waves we compute their density of
%states (DOS) as a function of energy. In fig(\ref{figura_dos_mf})
%we plot the density of states for the two band model and for the
%${\bf k } \cdot {\bf p}$ model.  In the calculation of the DOS the
%wave vector of the spin waves is restricted to be in an sphere of
%radius $k_D$, chosen to contain precisely the number of spins in
%the crystal. The DOS  is peaked at an energy $E_0$ proportional to
%the Curie temperature obtained in the virtual crystal
%approximation using the transverse susceptibility evaluated at a
%hole polarization $\xi$, $T^{vca} _{C,\xi}$=$\frac{S}{3} E _0$.

\subsection{Effect of the disorder}

In this subsection we analyze the effect of the disorder on
collective magnetic  excitations of the $T$=0 ferromagnetic ground
state. To this end, we diagonalize the equations of motion,
Eq.(\ref{eq_mot}),  for different disorder realizations. We
place  the Mn ions randomly on a FCC lattice, and  consider the
interaction of each Mn with all its  neighbors  within a
distance  shorter than  six lattice units.  We
consider systems with more than, typically, 500 Mn spins,  and  use periodic
boundary conditions. In the presence of disorder, the collective
excitations can not be characterized by a wave vector and we, therefore,
analyze their density of states (DOS). In
Fig.(\ref{figura_dos_disorder}) we plot the DOS of the collective
excitations for the two band model and for the for the six-band
${\bf k } \cdot {\bf p}$ model, for different values of $a_0$. The
DOS is obtained by averaging  over different disorder
realizations.  For every disorder realization, we always obtain a
zero energy mode that corresponds to an uniform rotation of all
the Mn spins. This Goldstone mode reflects the symmetry of the Hamiltonian
Eq.(\ref{heis_pol}).
\begin{figure}
% Requires \usepackage{graphicx}
  \includegraphics[clip,width=8cm]{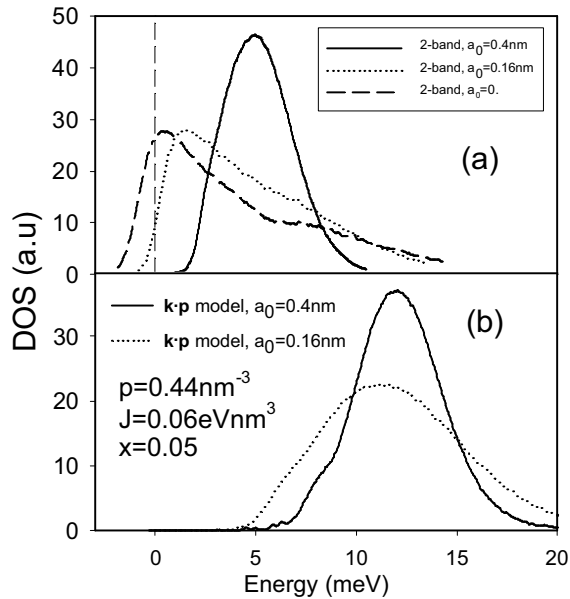}
 \caption{Density of states for low energy excitations
   in the two-band model and in the
  $\mathbf{k} \cdot \mathbf{p}$ model, obtained from the equation of motion method.
  The DOS is averaged over different disorder realizations.
  The parameters used are the same as in Fig.(\ref{figura_sw_mf}). }
 \label{figura_dos_disorder}
\end{figure}
In Fig.(\ref{figura_dos_disorder}a) we see that, in the  two band
model and for short  spatial extension of the spin interaction,
there are
 collective excitations with {\em negative} energy. This implies
that the  fully polarized GS  is unstable. This instability is due
to the long range oscillations of the interaction between the Mn
spins. For $a_0$=4\AA \   these oscillations  are
damped and  the fully polarized ferromagnetic GS is stable.

In Fig.\ref{figura_dos_disorder}(b), we plot the DOS for the six
band ${\bf k } \cdot {\bf p}$ model. Our results show  that the
fully polarized ferromagnetic GS is stable for any value of $a_0$
at the density of carriers studied.  This  stability is  the result of
the Fermi surface warping   produced by the heavy hole light
hole mixing in the ${\bf k } \cdot {\bf p}$ Hamiltonian. The holes
do not have a well defined 2$k_F$ anomaly in the
response functions.

This stability analysis obtained from the study of the $T$=0
ferromagnetic GS low energy magnetic  excitations is in agreement
with the results presented in the previous section. Note, however,
that only  the paramagnetic spin
susceptibility enters in the interaction
between Mn spins for the calculated Curie temperatures,.

 Recently Schlienman {\it et al.}\cite{Schliemann1,
Schliemann2} have computed the $T$=0 response function of DMS and
have suggested that, in general, the GS of DMS should be non
collinear. In their calculations they use a two band model, and
the instability is  the same that we find in
Fig.\ref{figura_dos_disorder}(a). From our calculations we
conclude that the instabilities  found in references
\cite{Schliemann1, Schliemann2} are just due to the model considered, and
disappear when a more realistic band structure model is used.

\section{Conclusions}
We have introduced a Heisenberg-like Hamiltonian,
Eq.(\ref{heis2}), for describing the magnetic properties of
GaMnAs. This model just requires the positions and orientations of
the Mn spins. The use of this model is justified because: i) the
energy of the system can be written as a sum of pair interactions,
ii) the interaction between two Mn spins is well described by
their scalar product and iii) the coupling constants of the
Heisenberg model depend basically on the distance between Mn spins, and
can be evaluated perturbatively.

As the electronic properties of the host semiconductor are
integrated into the coupling constants of the Heisenberg
Hamiltonian, it is possible to perfom MC simulations in systems
with a large number of Mn spins, and so to obtain the Curie temperature
of the DMS. We have compared the Monte Carlo, mean field and a
fluctuation corrected critical temperatures for different band
structure models and for different values of the exchange coupling
spatial extension $a_0$. The fluctuations effects are important
for large hole densities, being more relevant for smaller values
of $a_0$.

In the two band model, the existence of a well defined $2k_F$ anomaly
produces a long range interaction between Mn spins that
oscillates in sign with the distance. The fluctuations are
magnified by these oscillations which  can lead to the disappearance of the
ferromagnetic state from moderate to large values of the hole
density. We find that, for $a_0$ smaller than 1.6\AA \ and hole
densities larger than 0.44$nm^{-3}$ ,there is not a finite Curie
temperature. When the value of $a_0$ increases, the interaction
oscillations are damped and a finite Curie temperature appears.

In the ${\bf k} \cdot {\bf p}$ model, the  warping of the Fermi
surface softens the $2k_F$ anomaly in the response function and
the real space interaction oscillations  are almost suppressed.
Therefore, the effect of  fluctuations is weaker than in the two
band model. We find that, for $a_0$=4\AA,  the fluctuations effect
reduces the Curie temperature no more than 10$\%$. For smaller
values of  $a_0$,  the effect of the fluctuations is more severe,
with a reduction of  Curie temperatures by almost 50$\%$, for
$p$=0.66$nm^{-3}$. Nevertheless, we believe that a value of $a_0$
close to 4\AA \ is appropriated, as it corresponds to the first
neighbors distance in the FCC lattice. This is the minimum
separation between the Mn--$d$ orbitals and the GaAs-$p$ orbitals
that form the hole band of the semiconductor.

Finally, we have studied our model in the zero temperature limit.
We have computed the low energy collective magnetic excitations of
the $T$=0 ferromagnetic ground state.  In agreement with the Monte
Carlo results, we find that, for the two band model and  moderate
to large hole densities, the ferromagnetic ground state is no longer
stable. However,  when a more realistic six band ${\bf k} \cdot
{\bf p}$ model is used, the stability of the fully polarized
ferromagnetic ground state is recovered.

\centerline {\large{ \bf{Acknowledgements}}}

We are grateful to J.Fern\'andez Rossier and A.H.MacDonald for
helpful discussions. Financial support is acknowledged from Grants
No MAT2002-04429-C03-01 and MAT2002-04095-C02-01 (MCyT, Spain) and
Fundaci\'on Ram\'on Areces.

%\begin{figure}
%% Requires \usepackage{graphicx}
%  \includegraphics[clip,width=8cm]{figura_jrkp.eps} \\
%  \caption{Real space energy  coupling as function of $r$ for the $\mathbf{k}
%\cdot \mathbf{p}$
%  model with
%  the same parameters than in Fig.(\ref{figura_jqkp}).}
%\label{figura_jrkp}
%\end{figure}

%%%%%%%%%%%%%%%%%%%%%%%%%%%%%%%%%%%%%%%%%%%%%%%

%\bibliography{mia}

\begin{thebibliography}{26}
\expandafter\ifx\csname
natexlab\endcsname\relax\def\natexlab#1{#1}\fi
\expandafter\ifx\csname bibnamefont\endcsname\relax
  \def\bibnamefont#1{#1}\fi
\expandafter\ifx\csname bibfnamefont\endcsname\relax
  \def\bibfnamefont#1{#1}\fi
\expandafter\ifx\csname citenamefont\endcsname\relax
  \def\citenamefont#1{#1}\fi
\expandafter\ifx\csname url\endcsname\relax
  \def\url#1{\texttt{#1}}\fi
\expandafter\ifx\csname
urlprefix\endcsname\relax\def\urlprefix{URL }\fi
\providecommand{\bibinfo}[2]{#2}
\providecommand{\eprint}[2][]{\url{#2}}

\bibitem[{\citenamefont{Matsukura et~al.}(1998)\citenamefont{Matsukura, Ohno,
  Shen, and Sugawara}}]{Matsukura}
\bibinfo{author}{\bibfnamefont{F.}~\bibnamefont{Matsukura}},
  \bibinfo{author}{\bibfnamefont{H.}~\bibnamefont{Ohno}},
  \bibinfo{author}{\bibfnamefont{A.}~\bibnamefont{Shen}}, \bibnamefont{and}
  \bibinfo{author}{\bibfnamefont{Y.}~\bibnamefont{Sugawara}},
  \bibinfo{journal}{Phys.\ Rev.\ B} \textbf{\bibinfo{volume}{57}},
  \bibinfo{pages}{R2037} (\bibinfo{year}{1998}).

\bibitem[{\citenamefont{Ohno}(1998)}]{Matsukura-bis}
\bibinfo{author}{\bibfnamefont{H.}~\bibnamefont{Ohno}},
  \bibinfo{journal}{Science} \textbf{\bibinfo{volume}{281}},
  \bibinfo{pages}{951} (\bibinfo{year}{1998}).

\bibitem[{\citenamefont{Ohno and Matsukura}(2001)}]{Ohno}
\bibinfo{author}{\bibfnamefont{H.}~\bibnamefont{Ohno}} \bibnamefont{and}
  \bibinfo{author}{\bibfnamefont{F.}~\bibnamefont{Matsukura}},
  \bibinfo{journal}{Solid\ State\ Commun.} \textbf{\bibinfo{volume}{117}},
  \bibinfo{pages}{179} (\bibinfo{year}{2001}).

\bibitem[{\citenamefont{T.Dietl}(1994)}]{Dietlbook}
\bibinfo{author}{\bibnamefont{T.Dietl}}, \emph{\bibinfo{title}{Diluted Magnetic
  Semiconductors}}, vol.~\bibinfo{volume}{3B} of
  \emph{\bibinfo{series}{Handbook of Semiconductors}}
  (\bibinfo{publisher}{North-Holland}, \bibinfo{address}{New York},
  \bibinfo{year}{1994}).

\bibitem[{\citenamefont{Dietl et~al.}(2000)\citenamefont{Dietl, Ohno,
  Matsukura, Cibert, and Ferrand}}]{Dietl}
\bibinfo{author}{\bibfnamefont{T.}~\bibnamefont{Dietl}},
  \bibinfo{author}{\bibfnamefont{H.}~\bibnamefont{Ohno}},
  \bibinfo{author}{\bibfnamefont{F.}~\bibnamefont{Matsukura}},
  \bibinfo{author}{\bibfnamefont{J.}~\bibnamefont{Cibert}}, \bibnamefont{and}
  \bibinfo{author}{\bibfnamefont{D.}~\bibnamefont{Ferrand}},
  \bibinfo{journal}{Science} \textbf{\bibinfo{volume}{287}},
  \bibinfo{pages}{1019} (\bibinfo{year}{2000}).

\bibitem[{\citenamefont{Jungwirth et~al.}(1999)\citenamefont{Jungwirth,
  Atkinson, Lee, and MacDonald}}]{Jungwirth}
\bibinfo{author}{\bibfnamefont{T.}~\bibnamefont{Jungwirth}},
  \bibinfo{author}{\bibfnamefont{W.~A.} \bibnamefont{Atkinson}},
  \bibinfo{author}{\bibfnamefont{B.~H.} \bibnamefont{Lee}}, \bibnamefont{and}
  \bibinfo{author}{\bibfnamefont{A.~H.} \bibnamefont{MacDonald}},
  \bibinfo{journal}{Phys.\ Rev.\ B} \textbf{\bibinfo{volume}{59}},
  \bibinfo{pages}{9818} (\bibinfo{year}{1999}).

\bibitem[{not()}]{nota1}
\bibinfo{note}{Along this paper we take $g \mu _B =1 $ in the definition of the
  magnetic susceptibility.}

\bibitem[{\citenamefont{S.J.Potashnik et~al.}(2002)\citenamefont{S.J.Potashnik,
  K.C.Ku, R.Mahendiran, S.H.Chun, R.F.Wang, N.Samarth, and
  P.Schiffer}}]{Potashnik1}
\bibinfo{author}{\bibnamefont{S.J.Potashnik}},
  \bibinfo{author}{\bibnamefont{K.C.Ku}},
  \bibinfo{author}{\bibnamefont{R.Mahendiran}},
  \bibinfo{author}{\bibnamefont{S.H.Chun}},
  \bibinfo{author}{\bibnamefont{R.F.Wang}},
  \bibinfo{author}{\bibnamefont{N.Samarth}}, \bibnamefont{and}
  \bibinfo{author}{\bibnamefont{P.Schiffer}}, \bibinfo{journal}{Phys.\ Phys.\
  B} \textbf{\bibinfo{volume}{66}}, \bibinfo{pages}{012408}
  (\bibinfo{year}{2002}).

\bibitem[{\citenamefont{Edmonds et~al.}(2002)\citenamefont{Edmonds, Wang,
  Campion, Neumann, Farley, Gallagher, and Foxon}}]{Edmonds}
\bibinfo{author}{\bibfnamefont{K.~W.} \bibnamefont{Edmonds}},
  \bibinfo{author}{\bibfnamefont{K.~Y.} \bibnamefont{Wang}},
  \bibinfo{author}{\bibfnamefont{R.~P.} \bibnamefont{Campion}},
  \bibinfo{author}{\bibfnamefont{A.~C.} \bibnamefont{Neumann}},
  \bibinfo{author}{\bibfnamefont{N.~R.~S.} \bibnamefont{Farley}},
  \bibinfo{author}{\bibfnamefont{B.~L.} \bibnamefont{Gallagher}},
  \bibnamefont{and} \bibinfo{author}{\bibfnamefont{C.~T.} \bibnamefont{Foxon}},
  \bibinfo{journal}{Appl.\ Phys.\ Lett.} \textbf{\bibinfo{volume}{81}},
  \bibinfo{pages}{4991} (\bibinfo{year}{2002}).

\bibitem[{\citenamefont{Schliemann et~al.}(2001)\citenamefont{Schliemann,
  Konig, and MacDonald}}]{Schliemann}
\bibinfo{author}{\bibfnamefont{J.}~\bibnamefont{Schliemann}},
  \bibinfo{author}{\bibfnamefont{J.}~\bibnamefont{Konig}}, \bibnamefont{and}
  \bibinfo{author}{\bibfnamefont{A.~H.} \bibnamefont{MacDonald}},
  \bibinfo{journal}{Phys.\ Rev.\ B} \textbf{\bibinfo{volume}{64}},
  \bibinfo{pages}{165201} (\bibinfo{year}{2001}).

\bibitem[{\citenamefont{Alvarez et~al.}(2002)\citenamefont{Alvarez, Mayr, and
  Dagotto}}]{Dagotto-g}
\bibinfo{author}{\bibfnamefont{G.}~\bibnamefont{Alvarez}},
  \bibinfo{author}{\bibfnamefont{M.}~\bibnamefont{Mayr}}, \bibnamefont{and}
  \bibinfo{author}{\bibfnamefont{E.}~\bibnamefont{Dagotto}},
  \bibinfo{journal}{Phys.\ Rev.\ Lett} \textbf{\bibinfo{volume}{89}},
  \bibinfo{pages}{277202} (\bibinfo{year}{2002}).

\bibitem[{\citenamefont{Calder\'on et~al.}(2002)\citenamefont{Calder\'on,
  G.G\'omez-Santos, and Brey}}]{mjcdms}
\bibinfo{author}{\bibfnamefont{M.~J.} \bibnamefont{Calder\'on}},
  \bibinfo{author}{\bibnamefont{G.G\'omez-Santos}}, \bibnamefont{and}
  \bibinfo{author}{\bibfnamefont{L.}~\bibnamefont{Brey}},
  \bibinfo{journal}{Phys.\ Rev.\ B} \textbf{\bibinfo{volume}{66}},
  \bibinfo{pages}{075218} (\bibinfo{year}{2002}).

\bibitem[{\citenamefont{Berciu and Bhatt}(2000)}]{Berciu}
\bibinfo{author}{\bibfnamefont{M.}~\bibnamefont{Berciu}} \bibnamefont{and}
  \bibinfo{author}{\bibfnamefont{R.~N.} \bibnamefont{Bhatt}},
  \bibinfo{journal}{Phys.\ Rev.\ Lett.} \textbf{\bibinfo{volume}{87}},
  \bibinfo{pages}{7203} (\bibinfo{year}{2000}).

\bibitem[{\citenamefont{Schliemann and MacDonald}(2002)}]{Schliemann1}
\bibinfo{author}{\bibfnamefont{J.}~\bibnamefont{Schliemann}} \bibnamefont{and}
  \bibinfo{author}{\bibfnamefont{A.~H.} \bibnamefont{MacDonald}},
  \bibinfo{journal}{Phys.\ Rev.\ Lett.} \textbf{\bibinfo{volume}{88}},
  \bibinfo{pages}{137201} (\bibinfo{year}{2002}).

\bibitem[{\citenamefont{Schliemann}(2003)}]{Schliemann2}
\bibinfo{author}{\bibfnamefont{J.}~\bibnamefont{Schliemann}},
  \bibinfo{journal}{Phys.\ Rev.\ B} \textbf{\bibinfo{volume}{67}},
  \bibinfo{pages}{045202} (\bibinfo{year}{2003}).

\bibitem[{\citenamefont{G.Zar\'and and B.Jank\'o}(2002)}]{Zarand}
\bibinfo{author}{\bibnamefont{G.Zar\'and}} \bibnamefont{and}
  \bibinfo{author}{\bibnamefont{B.Jank\'o}}, \bibinfo{journal}{Phys.\ Rev.\
  Lett.} \textbf{\bibinfo{volume}{89}}, \bibinfo{pages}{047201}
  (\bibinfo{year}{2002}).

\bibitem[{\citenamefont{D.J.Priour and Sarma}()}]{Priour}
\bibinfo{author}{\bibfnamefont{E.}~\bibnamefont{D.J.Priour}} \bibnamefont{and}
  \bibinfo{author}{\bibfnamefont{S.}~\bibnamefont{Sarma}},
  \eprint{cond-mat/0305413}.

\bibitem[{\citenamefont{Okabayashi et~al.}(1998)\citenamefont{Okabayashi,
  Kimura, Rader, Mizokawa, Fujimori, Hayashi, and Tanaka}}]{Okabayashi}
\bibinfo{author}{\bibfnamefont{J.}~\bibnamefont{Okabayashi}},
  \bibinfo{author}{\bibfnamefont{A.}~\bibnamefont{Kimura}},
  \bibinfo{author}{\bibfnamefont{O.}~\bibnamefont{Rader}},
  \bibinfo{author}{\bibfnamefont{T.}~\bibnamefont{Mizokawa}},
  \bibinfo{author}{\bibfnamefont{A.}~\bibnamefont{Fujimori}},
  \bibinfo{author}{\bibfnamefont{T.}~\bibnamefont{Hayashi}}, \bibnamefont{and}
  \bibinfo{author}{\bibfnamefont{M.}~\bibnamefont{Tanaka}},
  \bibinfo{journal}{Phys.Rev.B} \textbf{\bibinfo{volume}{58}},
  \bibinfo{pages}{R4211} (\bibinfo{year}{1998}).

\bibitem[{\citenamefont{M.~Albolfath and MacDonald}(2001)}]{Albolfath}
\bibinfo{author}{\bibfnamefont{J.~B.} \bibnamefont{M.~Albolfath},
  \bibfnamefont{T.~Jungwirth}} \bibnamefont{and}
  \bibinfo{author}{\bibfnamefont{A.~H.} \bibnamefont{MacDonald}},
  \bibinfo{journal}{Phys.\ Rev.\ B} \textbf{\bibinfo{volume}{63}},
  \bibinfo{pages}{054418} (\bibinfo{year}{2001}).

\bibitem[{\citenamefont{Timm et~al.}(2002)\citenamefont{Timm, Schäfer, and von
  Oppen}}]{Timm}
\bibinfo{author}{\bibfnamefont{C.}~\bibnamefont{Timm}},
  \bibinfo{author}{\bibfnamefont{F.}~\bibnamefont{Schäfer}}, \bibnamefont{and}
  \bibinfo{author}{\bibfnamefont{F.}~\bibnamefont{von Oppen}},
  \bibinfo{journal}{Phys.\ Rev.\ Lett.} \textbf{\bibinfo{volume}{89}},
  \bibinfo{pages}{137201} (\bibinfo{year}{2002}).

\bibitem[{\citenamefont{P.Y.Yu and M.Cardona}(1996)}]{Cardona}
\bibinfo{author}{\bibnamefont{P.Y.Yu}} \bibnamefont{and}
  \bibinfo{author}{\bibnamefont{M.Cardona}}, in
  \emph{\bibinfo{booktitle}{Fundamentals of Semiconductors.}}
  (\bibinfo{publisher}{Springer-Verlag}, \bibinfo{address}{New York},
  \bibinfo{year}{1996}).

\bibitem[{\citenamefont{A.Baldereschi and N.O.Lipari}(1973)}]{Baldereschi}
\bibinfo{author}{\bibnamefont{A.Baldereschi}} \bibnamefont{and}
  \bibinfo{author}{\bibnamefont{N.O.Lipari}}, \bibinfo{journal}{Phys.\ Rev.\ B}
  \textbf{\bibinfo{volume}{8}}, \bibinfo{pages}{2697} (\bibinfo{year}{1973}).

\bibitem[{\citenamefont{Cardy}(1996)}]{cardy96}
\bibinfo{author}{\bibfnamefont{J.}~\bibnamefont{Cardy}},
  \emph{\bibinfo{title}{Scaling and Renormalization in Statistical Physics}}
  (\bibinfo{publisher}{Cambridge University Press},
  \bibinfo{address}{Cambridge}, \bibinfo{year}{1996}).

\bibitem[{\citenamefont{hot Lau and Dasgupta}(1989)}]{Lau}
\bibinfo{author}{\bibfnamefont{M.}~\bibnamefont{hot Lau}} \bibnamefont{and}
  \bibinfo{author}{\bibfnamefont{C.}~\bibnamefont{Dasgupta}},
  \bibinfo{journal}{Phys.\ Rev.\ B} \textbf{\bibinfo{volume}{39}},
  \bibinfo{pages}{7212} (\bibinfo{year}{1989}).

\bibitem[{\citenamefont{D.Pines and P.Nozieres}(1966)}]{Pines_Nozieres}
\bibinfo{author}{\bibnamefont{D.Pines}} \bibnamefont{and}
  \bibinfo{author}{\bibnamefont{P.Nozieres}}, in \emph{\bibinfo{booktitle}{The
  Theory of Quantum Liquids.}} (\bibinfo{publisher}{W.A.Benjamin, Inc.},
  \bibinfo{address}{New York}, \bibinfo{year}{1966}).

\bibitem[{\citenamefont{Konig et~al.}(2000)\citenamefont{Konig, Lin, and
  MacDonald}}]{Konig}
\bibinfo{author}{\bibfnamefont{J.}~\bibnamefont{Konig}},
  \bibinfo{author}{\bibfnamefont{H.~H.} \bibnamefont{Lin}}, \bibnamefont{and}
  \bibinfo{author}{\bibfnamefont{A.~H.} \bibnamefont{MacDonald}},
  \bibinfo{journal}{Phys.\ Rev.\ Lett.} \textbf{\bibinfo{volume}{84}},
  \bibinfo{pages}{5628} (\bibinfo{year}{2000}).

\end{thebibliography}

\end{document}